\documentclass[nonacm]{acmart} 

\usepackage{graphicx}
\usepackage{comment}
\usepackage{xspace}
\usepackage{color}
\usepackage{url}
\usepackage{csquotes}
\usepackage{hyperref}
\usepackage{cleveref}
\usepackage{caption}
\usepackage{outlines}

\usepackage[T1]{fontenc} 
\usepackage{anyfontsize}

\newcommand{\change}[1]{{#1}} 
\newcommand\bigoh{{\mathcal O}\xspace}

\newcommand{\sebi}{SeqBio\xspace}
\newcommand{\wca}{worst-case analysis\xspace}
\newcommand{\kmer}{$k$-mer\xspace}
\newcommand{\kmers}{$k$-mers\xspace}

\AtBeginDocument{%
  \providecommand\BibTeX{{%
    \normalfont B\kern-0.5em{\scshape i\kern-0.25em b}\kern-0.8em\TeX}}}

\setcopyright{acmcopyright}
\copyrightyear{2022}
\acmYear{2022}
\acmYear{}
\acmDOI{}
\acmVolume{}
\acmNumber{}

\begin{document}
\title{Theoretical analysis of edit distance algorithms: an applied perspective}

\author{Paul Medvedev}
\affiliation{%
\institution{The Pennsylvania State University}
  \city{University Park}
  \country{USA}}
  \email{pzm11@psu.edu}

\begin{abstract} 
Given its status as a classic problem and its importance to both theoreticians and practitioners, edit distance provides  an excellent lens through which to understand how the theoretical analysis of algorithms impacts practical implementations.  From an applied perspective, the goals of theoretical analysis are to predict the empirical performance of an algorithm and to serve as a yardstick to design novel algorithms that perform well in practice [Roughgarden 2019].  In this paper, we systematically survey the types of theoretical analysis techniques that have been applied to edit distance and evaluate the extent to which each one has achieved these two goals.  These techniques include traditional worst-case analysis, worst-case analysis parametrized by edit distance or entropy or compressibility, average-case analysis, semi-random models, and advice-based models.  We find that the track record is mixed.  On one hand, two algorithms widely used in practice have been born out of theoretical analysis and their empirical performance is captured well by theoretical predictions.  On the other hand, all the algorithms developed using theoretical analysis as a yardstick since then have not had any practical relevance.  We conclude by discussing the remaining open problems and how they can be tackled.
\end{abstract} 

\maketitle

\section{Introduction}

Edit distance, a classical problem in computer science, has received ongoing attention from both practitioners and theoreticians.
Given two strings $A$ and $B$, the edit distance is the minimum number of substitutions, insertions, and deletions needed to transform $A$ into $B$.
For example, the edit distance between $apfleee$ and $rapleet$ is 3: 
$apfleee \overset{\text{ins}}\longrightarrow rapfleee \overset{\text{del}}\longrightarrow rapleee \overset{\text{sub}}\rightarrow rapleet.$
The edit distance problem is widely known, as it is often taught as part of the undergraduate curriculum as a way to illustrate two-dimensional dynamic programming.
Theoreticians have studied the problem starting from as early as 1966~\cite{levenshtein1966binary} and 1974~\cite{wagner1974string},
but it very much remains an active topic of research today~\cite[e.g.][]{boroujeni2020improved}.
Simultaneously, bioinformatics practitioners continue to actively develop~\cite{edlib,parasail,seqan} and 
apply~\cite{racon,Maretty2017,Zook2019,isoncorrect,Sarkar2018,Brinda2017}
fast edit distance solvers ubiquitously.
Given its status as a classic problem and its importance to both theoreticians and practitioners, 
edit distance provides a unique opportunity to study the interaction between theory and practice.

Theoreticians develop abstract algorithms that have superior theoretical performance;
practitioners develop implemented algorithms that have superior empirical performance. 
In an ideal world, practitioners would implement and analyze the empirical performance of the abstract algorithms developed by theoreticians, while
theoreticians would analyze the theoretical performance of the implemented algorithms developed by practitioners.
In the real world, there is often a wide gap between the practical and theoretical communities; 
understanding how to close this gap is critical to making theoretical computer science more relevant to applications.
The edit distance problem is then an excellent lens through which to understand how the theoretical analysis of algorithms impacts practical implementations.

There are many ways to approach the practice/theory gap in a problem like edit distance.
We take one that is systematic and focused on the way theoreticians analyse edit distance algorithms, rather than on the algorithms themselves.
From a practical perspective, theoretical analysis has two goals~\cite{Roughgarden2019-dz}.
The first goal is to {\em predict} the empirical performance of an algorithm, either in an absolute sense or relative to other algorithms.
The second goal is to be a yardstick that drives the {\em design} of novel algorithms that perform well in practice.
In this paper, we will systematically survey the types of theoretical analysis techniques that have been applied to edit distance and evaluate the extent to which each one has achieved these two stated goals.



To focus our presentation, we consider only the most simple version of the edit distance problem,
where both strings have equal length $n$ and are over a constant sized alphabet; moreover, 
the algorithm only needs to return the edit distance and not the sequence of edits that achieve it.
We also focus only on the runtime analysis as opposed to the memory use.
We start by summarizing the state-of-the-art practical implementations (\Cref{sec:practice}).
We then go through the various types of theoretical analysis that has been applied to the edit distance problem:
traditional worst-case analysis (\Cref{sec:traditional}), 
worst-case analysis parametrized by the edit distance (\Cref{sec:parametrized}), 
worst-case analysis parametrized by entropy and compressibility (\Cref{sec:entropy}), 
average-case analysis (\Cref{sec:average}),
semi-random models (\Cref{sec:semi-random}),
and advice-based models (\Cref{sec:advice}).
For each technique, we evaluate the extent to which it has achieved the prediction and design goals of theoretical analysis on the edit distance computation problem.
We will not assume any knowledge of biology or any knowledge of computer science beyond an undergraduate level.
We then conclude with a discussion of open problems and their potential solutions.

\section{State-of-the-art implementations and algorithms}\label{sec:practice}
In this section we will briefly outline the algorithms that are used in state-of-the-art implementations as well as highlight their empirical performance.
The classical algorithm taught in many Algorithms courses is called Needleman-Wunsch~\cite{wagner1974string}.
It builds a two-dimensional matrix $D$ where the value at $D[i,j]$ is the edit distance between the $i$-long prefix of $A$ and the $j$-long prefix of $B$.
This matrix can be computed in the standard dynamic programming manner using the recurrence $D[i,j] = \min (D[i-1,j] + 1, D[i, j -1] + 1, D[i,j] + I_{i,j})$, 
where $I_{i,j}$ is 0 if the $i^\text{th}$ character of $A$ is equal to the $j^\text{th}$ character of $B$ and 1 otherwise.
The edit distance is then the value at $D[n,n]$ \change{(recall that $n$ is the length of the strings)}.
\change{Doubling banded alignment~\cite{Ukkonen1985-li} is a modification of Needleman-Wunsch whose main idea is to reduce the number of cells of $D$ that need to be computed. 
It uses the idea that if one only computes the values of $D[i,j]$  within a diagonal band of width $d$ (i.e. when $|i-j| \leq d/2$), 
then by checking if $D[n,m] \leq d$ one can either determine the edit distance if it is at most $d$ or determine that the edit distance is greater than $d$. 
Using this idea, it runs the checking algorithm repeatedly by doubling the value of $d$ until the edit distance is found.
}
The Myers' bit-parallel technique~\cite{myers1999fast}
is a hardware optimization of Needleman-Wunsch that encodes the dynamic programming matrix into bitvectors 
and then rewrites the recurrences in terms of word-sized bitvector operations.
These three algorithms/techniques comprise the core of all state-of-the-art widely-used implementations today.

There are at least three broadly used software libraries/tools that implement edit distance computation~\cite{edlib,seqan,parasail}.
Edlib~\cite{edlib} is optimized specifically for the edit distance problem, while SeqAn~\cite{seqan} and Parasail~\cite{parasail} are designed for more general alignment problems but support edit distance as a special case.
Edlib and SeqAn both implement the banded alignment algorithm using Myers' bit-parallel technique~\cite{myers1999fast}.
Myers' technique does not change the asymptotic runtime but gives a significant constant speedup in practice.
Parasail~\cite{parasail} implements the Needleman-Wunsch algorithm 
using high-performance computing techniques.
These include both task-level parallelism (i.e. multi-threading) 
and instruction-level parallelism (i.e. SIMD vectorized instructions).
Parasail's code is also customized during compilation for the instruction set of the host architecture. 
Another implementation, BGSA~\cite{bgsa}, implements the Needleman-Wunsh algorithm
with the Myers' bit-parallel technique but supports multi-core, task-level, and instruction-level parallelism 
for batch execution

There are also approaches to speed-up edit distance by using specialized hardware, such as GPUs, FPGAs, or even custom-designed processors (for references, see the introduction in~\cite{shouji}).
These result in orders-of-magnitude constant time speedups over their CPU counterparts in practice. 
However, until there is more wide-spread availability and integration of such
specialized hardware in bionformatics compute infrastructures,
these tools are unlikely to be widely used. 

How well do the widely used implementations perform?
On two sequences of 1 million nucleotides each, one of the fastest implementations (edlib)
takes 1.1 seconds for sequences with edit distance of $0.01n$ and 30 seconds for sequences with edit distance of $0.40n$, on a single core server~\cite{edlib}.
For sequences of 100,000 nucleotides each, the runtimes are $0.01$s and $0.40$s, respectively.
As we will later see, this corresponds to the theoretical prediction that the runtime deteriorates with increasing edit distance.
For many applications, these runtimes are good enough.
However, edit distance computation remains a bottleneck for applications that use it as a subroutine to make thousands or millions of comparisons.
(e.g. comparing long reads against each other~\cite{isoncorrect}).

\section{Traditional worst-case analysis}\label{sec:traditional}
The most common way to analyze running-time, taught in undergraduate computer science classes, is {\em traditional \wca}.
For example, it says that the classical merge sort algorithm runs in $\bigoh(n\log n)$ worst-case time, which formally means
that there exists a constant $c$ such that for any large-enough input of $n$ elements, merge sort takes at most $cn\log n$ time.
Has traditional worst-case analysis led to the design of edit distance algorithms that perform well in practice?
There are two candidates.
The first is the classical Needleman-Wunsch algorithm;
it was originally described in~\cite{wagner1974string}, 
which gave the dynamic programming recurrence and proved that the runtime is $\Theta(n^2)$.
Their algorithm modified an earlier dynamic programming algorithm~\cite{nw} whose run time 
was $\Theta(n^3)$.\footnote{As a historical note, the algorithm presented in the paper by Needleman and Wunsch~\cite{nw} is not the algorithm we call ``Needleman-Wunsch'' today.  
The ``Needleman-Wunsch'' algorithm was actually described later in~\cite{wagner1974string}.}
It seems likely then that traditional worst-case analysis was a driving force behind the creation of the Needleman-Wunsch algorithm.
Moreover, the algorithm is a success in practice because many implemented algorithms,
including some of the fastest ones, are either modifications of it or use it as a subroutine.
This includes both banded alignment and the Myers' bit-parallel technique used by edlib, parasail, and SeqAn.
Thus, though Needleman-Wunsch is not the fastest algorithm in practice or in theory, 
it exemplifies how traditional worst-case analysis achieved the {\em design} goal.

%
%

The second candidate is the fastest known algorithm 
under traditional worst-case analysis ---
the Four-Russians speedup to Needleman-Wunsch~\cite{fourrussians,masek1980faster}.   
It takes $\Theta(n^2/\log^2 n)$ time (in a unit-cost RAM model). 
The algorithm was clearly designed to optimize the runtime under traditional worst-case analysis.
But how does it perform in practice?
Does the $\Theta(\log^2 n)$ speedup outweigh the additional constant factors due to higher algorithm and data structure complexity?
To answer this question, there have been
implementations and experimental evaluations of this algorithm~\cite{rejmon,kim2016space}.
The improvement over Needleman-Wunsch was a factor of about 5 for $n=2^{18}$~\cite{rejmon}, and,
extrapolating from~\cite{rejmon}, would not exceed 10 even for sequences of a billion characters.
Thus, in practice, the Four Russians algorithm is dominated by other algorithms that have $\Theta(n^2)$ run time but have better constant factors (e.g. Myers' bit-parallel algorithm)~\cite{rejmon}.
Moreover, the fastest implementations of edit distance today~\cite{edlib,parasail,seqan} do not implement the Four Russians algorithm, even as a subroutine, 
highlighting how exploiting the properties of the CPU (e.g. the bit-parallel or SIMD implementations) can bring constant speedups that in practice outperform asymptotic speedups.
We therefore conclude that in the case of the Four Russians algorithm, 
traditional worst-case analysis has led us astray into the design of an algorithm that 
is not practically useful. 
\footnote{
	\change{
While the algorithm itself was not practically useful, 
one could argue that it had an impact on practice because some of its ideas were later used by algorithms such as Myers' bit-parallel algorithm.
More generally, while we may deem an algorithm not practically useful, 
it may nevertheless have been an important stepping stone on the road to another practically useful algorithm.
}
}

Has traditional worst-case analysis been able to accurately predict the empirical performance of algorithms?
The analysis of Needleman-Wunsch actually shows that
$\Theta(n^2)$ time is taken for every input instance, not just in the worst-case.
The Four-Russians runtime analysis is similar
in that it also holds for all inputs, not just worst-case ones.
Therefore, the predictions of traditional worst-case analysis 
accurately reflect these algorithms' runtime on real data.


Can traditional worst-case analysis lead to the design of new algorithms that perform well in practice?
A famous recent result states that under the strong exponential time hypothesis, there cannot be an $O(n^{2-\delta})$ algorithm, for any $\delta>0$~\cite{Backurs2018-ru}. 
Such an algorithm is called {\em strongly sub-quadratic}.
There are other ``barrier'' results of this type which we will not elaborate on~\cite{abboud2016simulating,abboud2015tight,bringmann2015quadratic}.
These results make it unlikely that better algorithms can be designed using traditional worst-case analysis as a guide.
However, they say nothing about the existence of provably and substantially better algorithms, as long as they are analyzed using a different technique than traditional worst-case analysis.

\section{Worst-case analysis parametrized by the edit distance}\label{sec:parametrized}
One step away from traditional worst-case analysis is {\em parametrized worst-case analysis},
which is worst-case analysis in terms of properties of the input
besides just its size $n$. 
In the case of edit distance, the parameter that has proven most useful is the edit distance itself,
usually denoted by $k$ (note that $k\leq n$).
The most notable algorithm designed using $k$-parametrized worst-case analysis is
doubling banded alignment~\cite{Ukkonen1985-li}.
Parametrized analysis shows that it computes the edit distance in time $\Theta(kn)$, on all inputs.
There are several other 
algorithms parametrized by $k$~\cite{navarro2001guided,Gusfield1997-ay};
they achieve various tradeoffs between $k$ and $n$ and have some other differences outside the scope of this article.
Doubling banded alignment is the most notable of these because it is very simple to describe and implement 
(i.e. it does not use any complex data structures)
and forms a main component of one of the 
empirically fastest implementation today (i.e. edlib~\cite{edlib}).

$K$-parametrized analysis predicts several ways in which doubling banded alignment is theoretically an improvement over Needleman-Wunsch.
The first is that when $k = o(n)$, doubling banded alignment scales sub-quadratically with the input size, while Needleman-Wunsch does not.
The second is that the closer the sequences are (i.e. the smaller the edit distance), the smaller the runtime. 
Thus the algorithm captures a natural notion of complexity of the input and takes advantage of input that is less complex,
while Needleman-Wunsch does not. 
These two predictions are reflected empirically as well, since the runtime analyses are for all inputs and both algorithms do not hide any significant implementation constants.
Third, even when $k=\Theta(n)$, the empirical advantage of doubling banded alignment over Needleman-Wunsch can be significant~\cite{edlib}.



However, when $k=\Theta(n)$, the fact remains that the banded algorithm scales quadratically with the input size, both in theory and in practice. 
Unfortunately, this is the case for the following predominant application of edit distance.
Biological sequences evolve from each other via a mutation process, which, for the purposes of this discussion, can be thought of as 
mutating each position with some constant probability, independently for each position. 
The mutation probability is, for the most part, independent of $n$ and the edit distance is therefore proportional to $n$ with a constant called {\em divergence}.
For example, the sequence divergence in coding regions of genes between human and other species is
1-2\% for bonobo and about 19\% for mouse~\cite{Cooper2003-cc}. 
Thus, regardless of species, and even for species that are very close, the edit distance is still roughly a constant proportion of the sequence length.
This illustrates the importance of 
using theoretical analysis to design exact algorithms that scale sub-quadratically 
when $k = \Theta(n)$.

In summary, the $k$-parameterized worst-case analysis technique has been a tremendous success for edit distance.
It led directly to the design of the banded alignment algorithm, which is widely used in practice, 
and it is able to predict the empirical improvement of banded alignment over Needleman-Wunsch. 
However, it still did not produce an algorithm that scales strongly sub-quadratically with the input size for applications where $k=\Theta(n)$; moreover, it may not be able to do so in the future due to the barrier results against strongly sub-quadratic algorithms in the worst-case framework~\cite{Backurs2018-ru,abboud2016simulating,abboud2015tight,bringmann2015quadratic}.

\section{Worst-case analysis parametrized by entropy and compressibility}\label{sec:entropy}

Another way to parameterize the analysis of edit distance algorithms is by the entropy of the input, $h$.
Entropy is a value between 0 and 1 which measures the amount of order in the strings~\cite{navarro2016compact}. 
Strings containing short repetitive patters tend to have a lower entropy.
As an extreme case, the string of all $T$s has entropy 0 while a string generated uniformly at random has entropy close to $1$.
Intuitively, an edit distance algorithm could take advantage of the repetitive patterns to 
run faster on strings with lower entropy.
An algorithm following this intuition was developed 
by~\cite{crochemore2003subquadratic} and runs in time $\bigoh(hn^2/\log n)$ for most inputs.
When the input strings have low entropy, this is theoretically faster than Needleman-Wunsch.
Note that this runtime is not comparable with banded alignment, since the edit distance can be low for strings with high entropy, and vice versa.

A related notion are algorithms that compute the edit distance of $A$ and $B$ directly from their compressed representations.
The time to compress two strings is in most cases asymptotically negligible compared to the time it takes to compute the edit distance.
Hence, one could solve the edit distance computation problem by first compressing $A$ and $B$ and then running an edit distance algorithm on the compressed strings.
For example, 
there is an algorithm to compute edit distance between two run-length encoded~\cite{navarro2016compact} strings in $\bigoh(\ell n)$ time~\cite{makinen2003approximate,crochemore2003subquadratic,arbell2002edit}, where $\ell$ is the size of the encoded strings. 
For a more general class of compression algorithms called straight-line programs~\cite{hermelin2009unified}, 
there is an algorithm \cite{gawrychowski2012faster} that runs in $\bigoh(\ell n \sqrt{\log (n / \ell)})$ time.
In these cases, the algorithms are designed to optimize the runtime with respect to $\ell$-parametrized worst-case analysis.


Unfortunately, the above algorithms have not been broadly applied in practice.
A major reason is that a long DNA sequence, while exhibiting some repetitive patterns, 
still has fairly high entropy (e.g. $0.85$~\cite{Loewenstern1999}) and low compressibility.\footnote{If compressing multiple DNA sequences together, much better compressibility is possible. 
However, edit distance has not been typically performed against such collections.}
Thus, while worst-case analysis parametrized by entropy or compressibility 
has led to the design of novel algorithms, it has not achieved the design or prediction goals of theoretical analysis since these algorithms have not been useful in practice.

\section{Average-case analysis}\label{sec:average}
One of the pitfalls of traditional and parametric worst-case analysis is that it assumes that the inputs are chosen by an adversary who wants to make things as difficult as possible for the algorithm.
In practice, however, biological sequences are not chosen this way and may have much nicer properties.
\change{On such instances, the optimal path in the alignment matrix stays very close to the diagonal, even 
	much tighter than the band of width $d$ used by the doubling banded alignment algorithm.
	For example, a heuristic algorithm that stops the doubling banded alignment prematurely and
	reports the value of $D[n,n]$ would still likely report the correct edit distance. 
	Worst-case analysis cannot tell us either the probability of success 
	nor tell us how long to double to guarantee a high probability of success.
}

One theoretical analysis technique to alleviate the shortcoming of worst-case analysis is average-case analysis (sometimes called distributional analysis).
Here, the inputs are assumed to be drawn from some kind of distribution, and 
what is measured is the expected performance over this distribution; 
other alternatives include measuring the performance that can be achieved with high probability
and/or measuring the performance in terms of a trade-off with the probability of the algorithm achieving it.
Re-analyzing the performance of existing algorithms under this model would not be beneficial 
for most of the algorithms we have looked at so far (i.e. Needleman-Wunsch, Four Russians, and banded alignment) because their runtime holds for all inputs, not just worst-case ones.
However, if we have an input distribution in mind, we can design a new algorithm to work well under that model and validate it empirically.
In particular, researchers have aimed to design an algorithm with a strongly sub-quadratic expected runtime.



An obvious first attempt is to assume that each string of length $n$ 
is randomly drawn uniformly and independently from the universe of all strings of length $n$.
For each string, this corresponds to generating each character independently, following an identical categorical distribution for each position.
However, this model does not capture essential properties of real data and has not resulted in any useful algorithms or predictive running time analysis.
When the real input strings are evolutionary related, 
then the assumption that they are independent of each other is highly inaccurate.
Even in the case that the real input strings are not evolutionary related, 
the uniformity assumption remains unrealistic, since biological sequences have evolved to serve a function 
and thus exhibit non-random behavior.

A more accurate model uses an indel channel~\cite{ganesh20near}.
Here, the first string is chosen uniformly at random, while the second string is obtained by randomly mutating each nucleotide, with a probability at most a small constant.
The algorithm of~\cite{ganesh20near} runs in $\bigoh(n \log n)$ time and, with high probability over this input distribution, returns the correct edit distance.
This algorithm's runtime is significantly better than just strongly sub-quadratic.
However, we are not aware of any implementation, and, in order for this algorithm to be practical, it would need to be shown that it performs reasonably well even for real input which does not follow the model's distribution. 

\change{
A successful application of average case analysis is in the analysis of the ``furthest reaching'' algorithm, which was proposed in~\cite{Ukkonen1985-li,myers1986ano}.
This analysis combines average-case with parametrized analysis.
It runs in $O(nk)$ time in the worst case~\cite{Ukkonen1985-li}, but an average case analysis gives $O(n + k^2)$~\cite{myers1986ano}.
The model used in this analysis is similar to the one in the previous paragraph which describes the model of~\cite{ganesh20near}.
While this algorithm does not break the quadratic time barrier and is not used by state of the art edit distance algorithms, 
it has nevertheless been practically successful. 
In particular, it was the basis of an implementation of the ``diff'' Unix tool~\cite{myers1986ano} and serves as a foundation for generalizations of the edit distance problem (e.g. to affine gap penalties as in~\cite{MarcoSola2021}).
}

\section{Semi-random models}\label{sec:semi-random}
One can view average-case analysis and worst-case analysis as two extremes.
The main idea of more sophisticated {\em semi-random} models
is to achieve a middle ground between an adversary and randomness~\cite{andoni2012smoothed,kuszmaul,boroujeni2020improved}.
Here, the performance is measured as worst-case over the choices of the adversary and average-case over the random distribution.
The semi-random models for edit distance have only led to approximation, rather than exact, algorithms.
A $c$-approximation algorithm is one that returns a value at most a multiplicative factor of $c$ (called the approximation ratio) away from the edit distance. 
Approximation algorithms relax the requirement of finding the exact solution in exchange for better runtime. 
In order for an edit distance approximation algorithm to be relevant in practice, the approximation ratio has to be a constant very close to one (e.g. 1.01).
For example, even a 3-approximation algorithm would not be able to always distinguish two random DNA sequences 
(i.e. expected edit distance of roughly $0.53n$ (personal simulations, data not shown)) from a mouse and human sequence pair 
(i.e. edit distance of about $0.19n$~\cite{Cooper2003-cc}).
Thus, the usefulness of the models is predicated on their ability to achieve a tiny approximation ratio.


\subsection{Smoothed analysis}
Smoothed analysis 
is based on the idea that the worst-case inputs that are designed to fool algorithms 
are not very stable; i.e. if bad input is tweaked a little, 
the algorithm no longer performs poorly~\cite{Roughgarden2019-dz}.
Generally speaking in this model, an adversary first picks an input (i.e. a worst-case choice),
but then some small random noise is added to this input. 
The algorithm's performance on a particular adversarial input is defined
as the expected performance over the distribution of noisy inputs centered around 
the adversarial input.
The algorithm's overall performance is then defined as the worst-case performance over all
choices of the adversary's input.

Smoothed analysis was considered for edit distance in~\cite{andoni2012smoothed}.
In their model, an adversary chooses the two input strings and a longest common subsequence between them. 
Then, each position is randomly perturbed with a small probability $p$ 
(i.e. the nucleotide is replaced with another random one);
however, the perturbations are constrained so that the positions of the longest common subsequence are perturbed identically.
This model captures the idea that for two evolutionary related strings,
if we view their longest common subsequence as their ancestral sequence,
then all the nucleotides outside this common subsequence would have evolved 
somewhat independently of each other. 
Allowing the adversary to choose some worst-case values for them gives her too much power;
instead, some noise is added to them to make them more independent.

In~\cite{andoni2012smoothed}, the algorithms' approximation ratios 
are not precisely derived (i.e. big-Oh notation is used). 
There has also been another line of work focusing on fast approximation algorithms under traditional worst-case analysis with the best known approximation constant of 1680~\cite{Chakraborty2018-vi}.
In both cases, the algorithms achieve strongly sub-quadratic run time, something that 
could not be done by exact algorithms using worst-case analysis for $k = \Theta(n)$.
However, without a precise derivation of a tiny approximation ratio, 
or an implementation and validation on real data,
it is difficult to predict the applicability of these algorithms.

\subsection{Asymmetric {\em (p,B)}-pseudorandom model}


As we mentioned previously, a string chosen uniformly at random does not reflect a biologically evolved sequence.
More precisely,
a uniformly random string has the property that the probability that any two non-overlapping equal-length substrings
are identical decreases with their length.
In fact, once their length exceeds a certain critical threshold, this probability is, for all practical purposes, zero. 
This property is in contrast with biological sequences, 
which are often composed of long similar elements called repeats.
It is true that longer repeats tend to be less frequent and less similar,
but this decrease does not happen at the same rate as for random strings. 
This was captured in a more realistic model of randomness,  proposed in~\cite{kuszmaul}.
They say that a string is {\em $(p,B)$-pseudorandom} 
if the edit distance of any two disjoint $B$-long substrings is at least a fraction $p$ of their length.
This generalizes uniform randomness, i.e. 
a uniformly random string is $(\Omega(1), \bigoh(\log n))$-pseudorandom 
with high probability~\cite{kuszmaul}.
But by choosing $p$ and $B$ appropriately, 
we can more realistically match the repeat properties of a biological string.

The asymmetric $(p,B)$-pseudorandom model~\cite{kuszmaul} is to first choose a string at random from all $(p,B)$-pseudorandom strings and
then have an adversary choose the other string and modify some small portion of the pseudorandom string.
By allowing the adversary to choose one of the strings, 
this model allows the two string to be evolutionary related. 
Moreover, the additional power of the adversary to modify the pseudorandom string makes the model even more realistic, because a true biological sequence would usually have some substrings that break the $(p,B)$, even when the values of $p$ and $B$ are chosen to minimize these cases. 

This model is used in~\cite{kuszmaul} to design several new algorithms. 
The main algorithm has a runtime $\widetilde{\bigoh}(nB)$ (the $\widetilde{\bigoh}$ notation is similar to $\bigoh$ but ignores log factors), which is strongly sub-quadratic as long as $B$ is strongly sub-linear.
However, as in smoothed analysis, the algorithm is only an approximation algorithm and
the exact approximation ratio is not calculated.
Thus, whether or not this model leads to any practical algorithms remains to be seen.
However, the fact that it seems to intuitively better capture biological reality 
while lending itself to theoretical analysis is promising.

\subsection{The random model of Boroujeni et al.}
A recently proposed model~\cite{boroujeni2020improved} has 
the adversary first choose a seed string $s$ and then constructs $A$
by permuting $s$ uniformly at random.
After observing $A$ and the random permutation, the adversary then constructs $B$.
The algorithm given by \cite{boroujeni2020improved} is an approximation algorithm with an expected runtime is $\bigoh(n^{1.898})$, which is strongly sub-quadratic.
The approximation ratio is $1 + o(1)$, which is low enough to be of practical relevance. 

This analysis model has led to the design of an algorithm that brakes through 
a barrier of previous models, 
i.e. it achieves a low approximation ratio while maintaining sub-quadratic time.
However, the actual runtime improvement is only $n^{0.102}$, which is less than 9 
for inputs up to a billion nucleotides long.
It is unlikely that this improvement would justify the additional overhead of a more complex algorithm.
Nevertheless, the model can ultimately be successful if the runtime of the algorithm can be improved,
the implementation of the algorithm kept simple, and the usefulness of the model empirically validated.


In summary, none of the semi-random models have yet led to a better understanding of the performance of existing algorithms or to the design of algorithms that perform well in practice.
The proposed algorithms, at least in their current form, are not promising: the algorithms 
of~\cite{andoni2012smoothed,kuszmaul}
have impractical approximation ratios, while the algorithm of~\cite{boroujeni2020improved} improves the runtime by a factor that is too small to have an effect in practice. 
However, the models themselves are promising, and can ultimately be successful if the runtime of the algorithms can be improved, the complexity of the algorithms kept low, and 
the usefulness of the models empirically validated.
These ongoing efforts may eventually result in an algorithm that outperforms banded alignment, in practice, on inputs with $k=\Theta(n)$.

\section{Analyzing with advice}\label{sec:advice}
The biological problem is usually more general than the mathematical abstraction created for it.
Sometimes the algorithm has access to other information, not included in the problem definition, that can serve as advice.
In this case, one can both expand the problem definition and the theoretical analysis to incorporate this advice.
An advice-based analysis measures the algorithm runtime with respect to the amount of such advice used.
In the case of edit distance,~\cite{goldwasser2017complexity}
argue that, for an input instance $A$ and $B$, it is possible to have access to a collection of correlated 
instances.
Intuitively, a correlated instance is one whose sequence of edits in the shortest edit sequence is, 
with some high probability, similar to the one between $A$ and $B$.
They show that if an algorithm has access to $\bigoh(\log n)$ of such instances, 
it can find the edit distance between $A$ and $B$ in $\bigoh(n\log n)$ time, with high probability.

This approach has not been implemented but is promising in the sense that the runtime is not just sub-quadratic but nearly linear, 
and the algorithm is exact and seems easy to implement. 
Compared to the banded alignment algorithms, 
the $\bigoh(n\log n)$ algorithm is likely to have significant empirical speed improvements for moderately sized inputs even for small values of $k$.
Unfortunately, it is not clear how correlated instances can be obtained in practice and
whether they would be captured by the definition of~\cite{goldwasser2017complexity}. 

\section{Conclusion}
We have surveyed the various approaches to the theoretical analysis of edit distance algorithms,
focusing on whether these approaches have led to the {\em design} of algorithms that are fast in practice and to the theoretical {\em prediction} of the empirical runtimes of existing algorithms. 
%
%
We showed that the track record has been mixed.
On one hand, a few algorithms widely used in practice (Needleman-Wunsch, doubled banded alignment, \change{furthest reaching}) have been born out of theoretical analysis and their empirical performance is captured well by theoretical predictions. 
On the other hand, very little of the algorithms developed using theoretical analysis as a yardstick since then have had any practical relevance.

From a practical perspective, a major open problem is to implement an algorithm with linear-like empirical scaling 
on inputs where the edit distance is linear in $n$.
Theoretical analysis has the potential to lead the way in achieving this goal.
Semi-random models are the most promising approach, due to the barrier results for worst-case analysis.
A reasonable model which might give a good balance between capturing reality and ease of analysis is one where 
$B$ is assumed to have evolved from $A$ via a mutation process (similar to~\cite{ganesh20near}). 
For the theoretical work to have practical relevance, however, it must be implemented and validated;
in particular, the algorithm's runtime and accuracy must be robust to data not drawn from the modeled distribution 
and the constant-time overhead of the algorithm must not override the asymptotic gains.
Unfortunately, current community incentives leave the implementation and validation of an algorithm as an afterthought that is rarely pursued.

In order to solve this open problem and, more generally, close the gap between theory and practice, 
implementation and validation cannot be treated as a separate step of the process.
We need multi-disciplinary teams that are able to interleave the theoretical analysis of algorithms with their implementation and validation. 
Allowing for a back-and-forth between practitioners and theoreticians during the development process 
can allow iterations over the theoretical model and practical heuristics that would otherwise be impossible.
Such teams will be able to use both empirical and theoretical performance as a yardstick and will be better able to 
develop algorithms whose empirical performance is not only superior but also accurately captured by theoretical analysis.

\change{
It is also essential to continue to study the relationship between theoretical analysis and practical implementations. 
One of the limitations of this study is that it only focuses on a single problem, 
making it difficult to draw more general conclusions. 
More studies such as this one could establish patterns and identify clear directions to closing the practice/theory gap.
In a recent paper~\cite{Medvedev2022limitations}, we applied the same lens more broadly (but also more anecdotally) 
to study the analysis of algorithms in sequencing bioinformatics,
attempting to establish patterns and draw conclusions for that domain.
Ultimately, the relationship between theory and practice must be understood not only from a technical angle but also from a social science and philosophy of science perspective; 
In other disciplines, this relationship is studied by philosophers of science (e.g. in education policy~\cite{joyce2020bridging}) and so a similar approach may be fruitful in computer science.
}

\change{
\paragraph{Acknowledgments:} 
This manuscript was inspired by the online videos for the 2014 class Beyond Worst-Case Analysis by Tim Roughgarden.  
This material is based upon work supported by the National Science Foundation under Grants No. 1439057, 1453527, and 1931531.
}

\bibliographystyle{plain}
\bibliography{tasba}

\end{document}